\documentclass[12pt]{article}

\usepackage{graphicx}
\usepackage{color}
\usepackage{amsmath,amssymb,amsfonts}
\usepackage{bbm}
\begin{document}
\author{Sven  Gnutzmann\thanks{sven.gnutzmann@nottingham.ac.uk} \hspace{0.05cm} and Stylianos Lois\\
School of Mathematical
  Sciences\\ University of Nottingham\\ Nottingham NG7 2RD, UK
}
\title{Remarks on nodal volume statistics\\ for regular and chaotic wave functions\\ in various dimensions} 

\maketitle
\begin{abstract}
  We discuss the statistical properties of the volume of the nodal set
  of wave function for two paradigmatic model systems which we
  consider in arbitrary dimension $s\ge 2$: the cuboid
  as a paradigm for a regular shape with separable wave functions,
  planar
  random waves as an established model for chaotic wave functions
  in irregular shapes. We give explicit results for the mean and
  variance of the nodal volume in arbitrary dimension, and for their
  limiting distribution. For the mean nodal volume we calculate
  the effect of the boundary of the cuboid where Dirichlet boundary
  conditions  reduce the nodal volume compared to the bulk. 
  Boundary effects for chaotic wave functions are calculated using 
  random waves which satisfy a Dirichlet boundary condition 
  on a hyperplane. \\
  We put forward several conjectures what properties of
  cuboids generalise to general
  regular shapes with separable wave functions and what
  properties of random waves can be expected for general
  irregular shapes. These universal features clearly distinct
  between the two cases.
\end{abstract}
\section{Introduction}
We consider real square-integrable eigenfunctions $\Phi(\mathbf{q})$
of the free stationary Schr\"odinger (or Helmholtz) equation 
\begin{equation}
  - \Delta_{\mathcal{M}}\Phi(\mathbf{q}) = E \Phi(\mathbf{q}) 
  \label{schrod}
\end{equation}
on an $s$-dimensional smooth connected compact
Riemannian manifold $\mathcal{M}$ with local coordinates
$\mathbf{q}\equiv(q^1,\dots,q^s)$ .  
Here, $\Delta_{\mathcal{M}}$ is
the Laplace-Beltrami operator on $\mathcal{M}$ and $E$ is an energy eigenvalue.  We have set the value of
the physical constant $\frac{\hbar^2}{2m}$ of Planck's constant
squared over twice the mass of the particle equal to one by
appropriate choice of units.\\
If $\mathcal{M}$ has a boundary we will impose Dirichlet
boundary conditions.\\
Compactness ensures a discrete and non-negative energy spectrum
that we arrange in ascending order as $0 \le E_1 < E_2 \le
\cdots \le E_N\le E_{N+1}\le \cdots$. 
The eigenfunction with eigenvalue $E_N$ will be denoted by
$\Phi_N(\mathbf{q})$ (for degenerate eigenvalues this requires a
choice of eigenbasis and order). \\
For a
given eigenfunction $\Phi_N(\mathbf{q})$ the nodal set 
\begin{equation}
\mathcal{N}[\Phi_N(\mathbf{q})]=
\Phi_N^{-1}(0) \backslash \partial M \subset\mathcal{M}
\label{nodalset}
\end{equation} 
consists of all \textit{interior} points on the
manifold where the eigenfunction vanishes.
For the real wave functions that are under consideration here the
nodal
set is a collection of hyper-surfaces.

More than 200 years ago Chladni \cite{Chladni}  visualised the
vibration modes of plates with sand that accumulates along nodal
lines.
He analysed in detail the geometric patterns formed by the nodal lines.
About 30 years later Sturm's oscillation theorem \cite{Sturm} 
that the $n$-th eigenfunction of a Sturm-Liouville
differential operator has $n-1$ nodal points may have been the first
rigorous mathematical result concerning the nodal set
of wave functions. Since this time the nodal set for wave functions
of various types has attracted the attention of many mathematicians
and physicists -- and many seminal results followed.

In the present work we will focus on the size of the nodal set
which we measure by its hyper-surface volume. We denote the
hyper-surface volume of the nodal set of the $N$-th eigenfunction by
$\mathcal{H}_N$ -- we will refer to it just as the \emph{nodal volume}.\\
With increasing energy the wavelength becomes smaller and so does the
typical distance between two nearby nodal surfaces.
One thus expects that typical $\mathcal{H}_N$ increase
with the energy $E_N$. Comparison of nodal 
volumes between eigenfunctions at very different
energies or of eigenfunctions on different manifolds is possible
through the dimensionless \emph{rescaled nodal volume}
\begin{equation}
  \sigma_N= \frac{\mathcal{H}_N}{\mathcal{V} \sqrt{E_N}}
  \label{definition_sigma}
\end{equation}
where $\mathcal{V}$ is the volume of the manifold $\mathcal{M}$.
Note that the rescaled nodal volume is only defined for positive 
energies $E_N>0$. In manifolds without boundary this excludes
the ground state with energy $E_1=0$ which has a vanishing
nodal set and thus $\mathcal{H}_1=0$.

In the mathematical literature the nodal volume has been a central
object and the main results are best summarised by Yau's
conjecture \cite{yau} that states that the rescaled nodal volumes are bounded
from below and above for all smooth compact manifolds. That is
\begin{equation}
  c_1 \le \sigma_N \le c_2
\end{equation}
for all $N\ge 2$ where the constants $c_1$ and $c_2$ only depend on the
manifold and the metric. For real analytic manifolds this is a classic
theorem by Donnelly and Fefferman \cite{donnelly}. For the smooth case 
lower bounds have been established \cite{bruning1,bruning2,colding}. 
The general proof of Yau's conjecture
remains a central problem in spectral theory -- we refer to the recent
survey by Zelditch for a more complete overview and additional 
results \cite{zelditch}.

In this work we will consider how the rescaled nodal volume
for a given manifold is distributed statistically. Such a 
statistical approach is well established for the number
of nodal domains \cite{BGS,Bogomolny} which revealed that
shapes with a chaotic ray dynamics have a universal distribution
which can clearly be distinguished from distributions for
shapes with a separable Laplacian and thus integrable
ray dynamics. For separable shapes these distributions can often be
calculated explicitly \cite{BGS,GL} and share some universal features.
Here we will consider nodal volumes for two models:
\begin{itemize} 
  \item[i.] The cuboid. This is a paradigm of a 
    regular shape with separable Laplacian and
    integrable ray dynamics;
  \item[ii.]
    The boundary adapted planar random wave model \cite{berry-nod}.
    In two dimensions statistical properties of nodal volumes
    and nodal densities
    have been discussed in detail for various types of random superposition of 
    eigenstates 
    \cite{berry-nod2,berry-nod3,dennis,oravecz,rudnick,wigman1,wigman-fluct}.
    The the boundary-adapted random wave model 
    is an extension of the standard Gaussian random wave model
    introduced by Berry \cite{berry-rw} who conjectured that
    eigenfunction statistics for chaotic systems follows the predictions 
    of the Gaussian random wave model in a semiclassical limit.
    The boundary adapted random wave model 
    is able to predict systematic corrections 
    near the  boundary -- for the two-dimensional case
    the effect of a boundary on nodal densities has been discussed in
    detail \cite{berry-nod,berry-nod2,berry-nod3}. We will add some
    results for higher dimensions and discuss implications for 
    manifolds with chaotic ray dynamics.
\end{itemize}
In section \ref{regular} we will give a full derivation of 
limiting distributions for the nodal volume based on Poisson
summation. Moreover we will consider boundary corrections
to the mean of the nodal volume.
In section \ref{irregular} we will summarise Berry's results
on the nodal volumes in boundary adapted random waves
and derive some extensions. The main result of this chapter is formulated
as a conjecture on the finite energy correction to the mean nodal volume in 
an irregular shape for arbitrary dimension.
Implications of our findings to more general regular and irregular
shapes will be discussed further in section
\ref{discussion}. 

\section{The regular case: nodal volume statistics for an 
  $s$-dimensional cuboid}
\label{regular}

Let $a_\ell$ ($\ell=1,\dots,=s$) be the side lengths of an $s$-dimensional
cuboid with volume $\mathcal{V}=\prod_{\ell=1}^s a_\ell$. 
Separation of variables leads to a unique basis of
(normalised) eigenfunctions
\begin{equation}
  \Phi_{\mathbf{n}}(\mathbf{q})= \frac{(2\pi)^{s/2}}{\mathcal{V}^{1/2}}
  \prod_{\ell=1}^s \sin\left( \frac{
      \pi n_\ell q_\ell}{a_\ell} \right)
\end{equation}
which are labelled by $s$ positive integers $n_\ell$
($\ell=1,\dots,s$). The corresponding energies are
\begin{equation}
  E_{\mathbf{n}}= \pi^2 \sum_{\ell=1}^s \frac{n_\ell^2}{a_\ell^2}
\end{equation}
and the rescaled nodal volumes are given by
\begin{equation}
  \sigma_{\mathbf{n}}= \frac{1}{\sqrt{E_{\mathbf{n}}}} \sum_{\ell=1}^s
    \frac{n_\ell-1}{a_\ell}\ .
\end{equation}
Our first goal will be to obtain the asymptotic mean value 
of $\sigma_{\mathbf{n}}$
in a spectral interval $E_{\mathbf{n}} \in [E, E+\Delta E]$
of width $\Delta E$ near the energy $E$
\begin{equation}
  \langle \sigma_{\mathbf{n}} \rangle_{[E,E+\Delta E]}
  = \frac{1}{N_{[E,E+\Delta E]}} \sum_{\mathbf{n} \in \mathbbm{N}^s}
  \sigma_{\mathbf{n}}\ \chi_{[E,E+\Delta E]}(E_{\mathbf{n}}) \ .
  \label{meansigma}
\end{equation}
Here $N_{[E,E+\Delta E]}$ is the number of eigenfunctions with energies
in the interval 
$E_{\mathbf{n}}\in [E,E+\Delta E]$ and $\chi_{[E,E+\Delta E]}(x)$ is the
characteristic function of this interval.

We will be interested in the asymptotic behaviour as $E \rightarrow
\infty$ such that $N_{[E,E+\Delta E]} \rightarrow \infty$ at the same time.
The choice $\Delta E= g E^{1/4}$ for some constant $g>0$ 
satisfies the above requirement and is sufficiently
small such that any systematic change
over the interval only leads to corrections which are much smaller
than the ones we will explicitly calculate. 
The standard tool for extracting asymptotic behaviour in
this setting is Poisson's summation formula. We will
apply it in the form
\begin{equation}
  \sum_{n=1}^\infty f(n) = \int_0^\infty f(n) dn -\frac{1}{2} f(0)+
  2 \sum_{M=1}^\infty \int_0^\infty f(n)\cos(2\pi Mn) dn
  \label{poisson}
\end{equation}
which is valid for sufficiently well-behaved functions $f(n)$
(such that all sums and integrals converge). 
Poisson summation allows us to find the asymptotic behaviour
of $N_{[E,E+\Delta E]}= N(E+\Delta E)- N(E)$
where
\begin{align}
   N(E)=& \sum_{\mathbf{n} \in \mathbbm{N}^s} \theta(E-E_{\mathbf{n}})\\
   = &N_{\mathrm{Weyl}}(E) + N_{\mathrm{osc}}(E)
\end{align}
is the spectral counting function.
In the second line we have written $N(E)$ as a sum of a smooth
monotonically increasing function $N_{\mathrm{Weyl}}(E)$ and an
oscillating function $N_{\mathrm{osc}}(E)$.
\begin{figure}[ht]
  \begin{center}
    \includegraphics[width=0.7\textwidth,height=0.45\textheight]{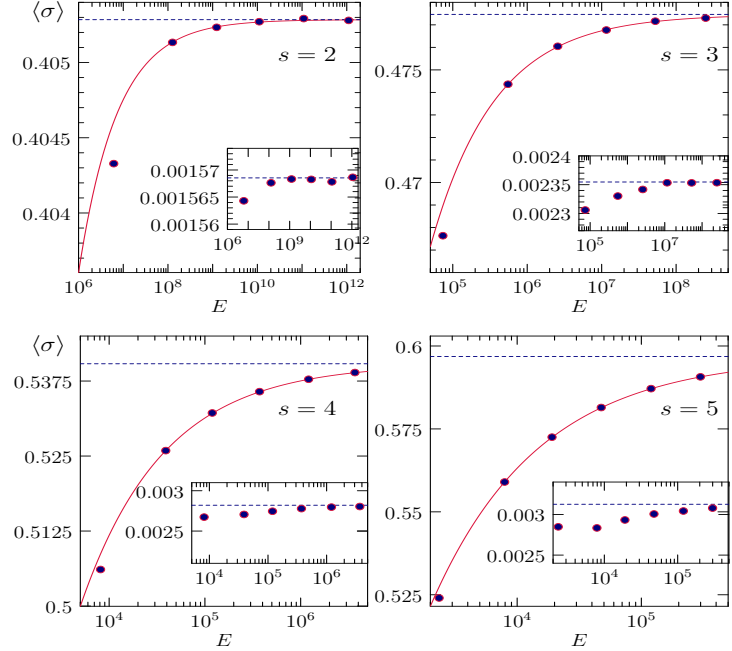}
    \caption{\label{fig_meanvar} Mean (main panels) and variance
      (insets) of the rescaled
      nodal volume as a function of energy. Data points are averages over
      $\approx 10^6$
      eigenstates in an interval around the given energy.}
  \end{center}
\end{figure}
Application of the Poisson summation formula reveals that $N(E)$
is
asymptotically dominated by the smooth part
\begin{equation}
  N_{\mathrm{Weyl}}(E)= \frac{\xi_s \mathcal{V}}{2^s\pi^s}  E^{s/2}
  - \frac{\xi_{s-1} \mathcal{S} }{2^{s+1} \pi^{s-1}} E^{(s-1)/2}
  + \mathit{O}(E^{(s-2)/2})
  \label{weyl}
\end{equation}
which is known as Weyl's law. 
Here 
$
  \xi_s=\frac{
    \pi^{s/2}}{\Gamma\left(\frac{s}{2}+1\right)}
$
is the volume of the
$s$-dimensional unit ball
and $\mathcal{S}=2 \mathcal{V} \sum_{\ell=1}^s\frac{1}{a_\ell}$
is the $s-1$ dimensional volume of the 
surface of the cuboid.\\
The first term in \eqref{weyl}
gives the leading growth of the number of states
with increasing energy. The second term is the leading correction --
a boundary effect as can be seen from the appearance of the
surface volume. 
Each oscillating contribution is of
order $\mathit{O}(E^{(s-1)/4})$ and thus asymptotically smaller than 
the boundary correction.\\
Weyls law implies the estimate
\begin{equation}
  N_{[E,E+\Delta E]}= \frac{\xi_s \mathcal{V}}{2^s \pi^s} E^{s/2} \frac{s\Delta E}{2E}
  \left(1 - \frac{(s-1) \pi \xi_{s-1} \mathcal{S}}{2 s \xi_s \mathcal{V} } E^{-1/2}
    +\mathit{O}\left(E^{-3/4}\right)\right) 
\end{equation}
which includes the effect of the boundary in the leading correction.
Here $\Delta E= g E^{1/4}$ has been used to give the order of further
corrections which
will be neglected in the sequel.\\
We may now apply Poisson summation to find the mean value of the
rescaled
nodal volume
\begin{align}
  \left\langle \sigma_{\mathbf{n}}
  \right\rangle_{ [E, E+\Delta E ] }
  =\frac{2 \xi_{s-1}}{\pi \xi_s}
    \left(
      1-
      \beta_s \frac{\mathcal{S}}{\mathcal{V}} E^{-1/2} +\mathit{O}(E^{-3/4})
    \right)
    \label{asymptotic_mean}
\end{align}
where
\begin{equation}
  \beta_s=
  \frac{(s-1) \pi \xi_{s-2}}{2s\xi_{s-1}}+\frac{ \pi
    \xi_s}{4\xi_{s-1}}-\frac{(s-1)\pi\xi_{s-1}}{2s\xi_s}\ .
\end{equation}
The three terms in the above expression for $\beta_s$ have different
origins in the asymptotic expansion of the Poisson sum.
The nodal volume $\sigma_{\mathbf{n}}$ contains terms proportional to $n_\ell-1$.
The terms proportional to $n_\ell$
dominate give the leading term in
\eqref{asymptotic_mean} and the first term of $\beta_s$.
The unit shift leads to the second term in $\beta_s$ and the third term
comes from the boundary correction to the spectral counting function.
Note that $\beta_s$ is a positive constant for any $s$ and that it
decays as the dimension grows.

Analogously one may calculate the variance which we only give to
leading order 
\begin{equation}
  \mathrm{Var}(\sigma_{\mathbf{n}})_{[E,E+\Delta E]}=
  \frac{1}{\pi^2}+
  \frac{4 (s-1) \xi_{s-2}}{s \pi^2 \xi_s}
  - \frac{4 \xi_{s-1}^2}{\pi^2 \xi_s^2}+
  \mathit{O}(E^{-1/2}) \ .
\end{equation}
Higher moments can be obtained in a similar way.
Alternatively one may just consider the limiting
distribution 
\begin{equation}
  P_s(\sigma)= \lim_{E\to \infty}\langle
  \delta(\sigma-\sigma_{\mathbf{n}})\rangle_{[E,E+\Delta E]} 
\end{equation}
for which one may derive the formal expression
\begin{equation}
  P_s(\sigma)=\frac{1}{s \xi_s}\int d^{s-1}\Omega\ 
  \delta\left(\sigma- \frac{\sum_{\ell=1}^s |e_\ell|}{\pi}\right)
  \label{limdist}
\end{equation}
\begin{figure}[htb]
  \begin{center}
    \includegraphics[width=0.7\textwidth,height=0.45\textheight]{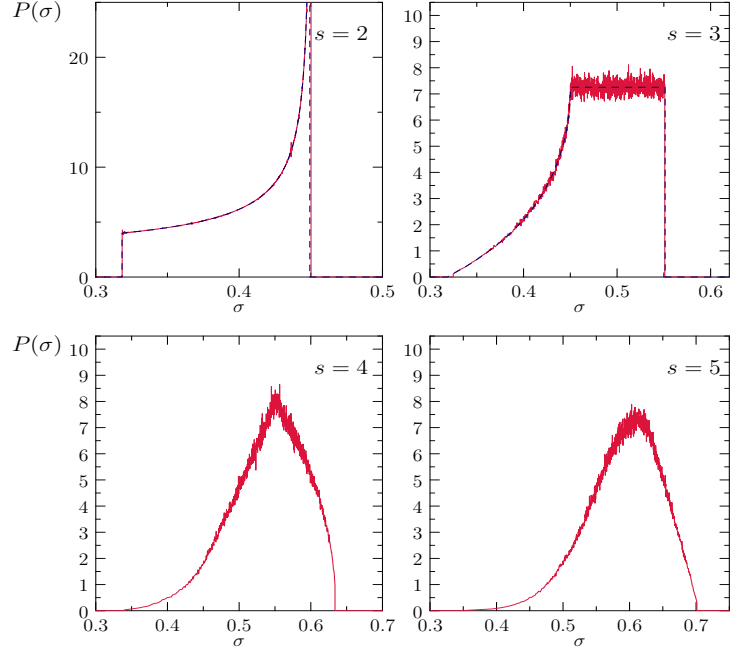}
    \caption{\label{fig_dist} Nodal volume distributions for an
      $s$-dimensional cuboid. The red curves show histograms obtained
      from $10^6$ highly excited eigenstates. For $s=2$ and $s=3$ the
      upper panels also show the limiting distributions \eqref{dist2}
      and \eqref{dist3} as a dashed line.}
  \end{center}
\end{figure}
where the integral is over the a unit $s-1$-dimensional sphere and
$\mathbf{e}=(e_1,\dots,e_s)$ is a point on the sphere
(i.e. $\mathbf{e}^2=1$).
It is straight forward to see that these limiting distributions vanish
outside the interval $\sigma\in \left[\frac{1}{\pi}, \frac{\sqrt{s}}{\pi} \right]$.
For low dimensions \eqref{limdist} can be given by direct
integration
\begin{align}
  P_2(\sigma)=&
  \begin{cases}
    \frac{4}{\sqrt{2-\pi^2 \sigma^2}}
    & \text{for $\sigma\in [1/\pi,\sqrt{2}/\pi]$,}\\[0.5cm]
    0 &\text{else;}
  \end{cases}\label{dist2}\\[0.5cm]
  P_3(\sigma)=&
  \begin{cases}
    \frac{4}{\sqrt{3}}
    \left(
      \frac{\pi}{2}  +
      \arctan\left(\frac{\pi\sigma}{\sqrt{6-3\pi^2\sigma^2}} \right)
      + \right. &\\
    \qquad\arctan\left(\frac{\pi\sigma-3\sqrt{2-\pi^2\sigma^2}}{\sqrt{6+6\pi\sigma\sqrt{2-\pi^2\sigma^2}}}
    \right) -&\\
    \qquad\left. 
      \arctan\left(\frac{\pi\sigma+3\sqrt{2-\pi^2\sigma^2}}{\sqrt{6-6\pi\sigma\sqrt{2-\pi^2\sigma^2}}}
      \right)
    \right)
    & \text{for $\sigma\in [1/\pi,\sqrt{2}/\pi]$,}\\[0.5cm]
    \frac{4 \pi}{\sqrt{3}} & \text{for $\sigma\in [\sqrt{2}/\pi,\sqrt{3}/\pi]$,}\\[0.5cm]
    0 &\text{else.}
  \end{cases}\label{dist3}
\end{align}
For larger dimensions direct integration of the expression
\eqref{limdist} may 
be performed with increasing effort. In Figure \ref{fig_dist} histograms
of these distributions for a finite energy interval are shown that
illustrate how the distribution changes with increasing dimension. 
\\
Let us conclude this section with a few observations about this
example which may be generalised to more general separable cases.
Apart from points of high symmetry separable wave functions all have
a similar local checker board structure. One may then expect that the qualitative behaviour of
the mean nodal volume (including the boundary effect), the variance
and the limiting distribution of nodal volumes will be very similar.
We conjecture that the following features of the limiting
distribution are universal
\begin{itemize}
  \item[i.] The limiting distribution has compact support
    $\sigma\in [\sigma_{\mathrm{min}},\sigma_{\mathrm{max}}]$ with
    $\sigma_\mathrm{min}>0$ (this is consistent with Yau's conjecture).
  \item[ii.] Near $\sigma_{\mathrm{min}}$ the behaviour is 
    $P(\sigma)= c_1 (\sigma-\sigma_{\mathrm{min}})^{s-2}  \ \theta(\sigma-\sigma_{\mathrm{min}})$.
  \item[iii.] Near $\sigma_{\mathrm{max}}$ the behaviour is 
    $P(\sigma)= c_2 (\sigma_{\mathrm{max}}-\sigma)^{(s-3)/2}  \
    \theta(\sigma_{\mathrm{max}}-\sigma)$ (see \cite{GL} for an
    analogous singularity in limiting distributions of nodal counts).
\end{itemize}

\section{The chaotic case: nodal volume statistics of boundary-adapted random waves}
\label{irregular}

Let us now  consider nodal volume statistics for $s$-dimensional
random waves as proposed by Berry \cite{berry-rw}. 
These are a model for wave
functions in chaotic billiards. In order to
account for boundary corrections to the mean nodal volume we use
boundary adapted random waves following closely
the analysis presented by Berry in \cite{berry-nod}, where the mean 
nodal volume
for random waves in $s=2$ dimensions which satisfy a Dirichlet
boundary condition along an infinite line was calculated 
(see also \cite{berry-nod2,berry-nod3,dennis}).
In the mathematical literature similar approaches have been used
to study statistical properties of nodal volumes on tori \cite{oravecz,rudnick}
and spheres \cite{wigman1,wigman-fluct}. In these cases random wave
models have been constructed in terms of random superpositions of degenerate
eigenfunctions and rigorous results on expected nodal volumes and on the 
fluctuations were obtained. The latter are largely consistent with the results
obtained by Berry for planar random waves for appropriate choices of 
eigenspaces.

Let us now construct
a Gaussian random wave model whose realisations are
solutions of the $s$-dimensional
free stationary Schr\"odinger equation \eqref{schrod} on 
the Euclidean space $\mathbbm{R}^s$
at energy $E=k^2$  
with a Dirichlet condition $\Phi=0$ on the hyperplane
$x_s=0$ (where $\mathbf{x}=(x_1,\dots,x_s)\in \mathbbm{R}^s$ are Cartesian 
coordinates). 
We will consider solutions in the half space $x_s>0$ and thus refer to
the plane $x_s=0$ as the boundary. 
It will be convenient to use `dimensionless' (rescaled) coordinates, 
so we define
$\mathbf{R}=(R_1,\ldots , R_s)$, where $R_i= kx_i$. 

Let us first define
the standard random waves (without any boundary conditions) by
\begin{equation}
  u(\mathbf{R})= \mathrm{Re} \sqrt{\frac{2}{N}} \sum_{j=1}^{N} e^{i \mathbf{R} \cdot \mathbf{n_j}  + i\phi_j}
  \label{rwm_iso}
\end{equation}
where $\mathbf{n_j}$ are uniformly distributed on a unit
$(s-1)$-sphere and the phases $\phi_j$ are uniformly distributed on
$[0,2\pi)$. The Gaussian random wave model is achieved in the formal
limit $N \to \infty$. The random waves \eqref{rwm_iso}
do not obey any boundary conditions.
Dirichlet boundary conditions at the boundary $R_s=0$ can be implemented in a
straight forward way by  anti-symmetrisation with respect to the boundary
\begin{equation}\label{anisrwm}
  \Phi(\mathbf{R})= \frac{1}{\sqrt{2}} (u(\mathbf{R})-u(\mathbf{\tilde R}))
\end{equation}
where $\mathbf{\tilde R}\equiv \mathbf{R}- 2 R_s \mathbf{e}_s$
where $\mathbf{e}_s=(0,\dots,0,1)$ is the unit vector in direction $x_s$. 
It is expected that the effect of the boundary at $x_s=0$ becomes
weaker when $x_s$ becomes larger.
We have chosen the normalization constant in \eqref{anisrwm}  
so that $\langle \Phi^2\rangle \rightarrow 1$ as $R_s\rightarrow \infty$
where $\langle \cdot \rangle$ refers to the average over random waves. 

For a given region $G$ in the half space $x_s>0$ ($R_s>0$) we may now write
the expected rescaled nodal volume as
\begin{equation}
  \begin{split}
    \left\langle \sigma_G \right\rangle =& 
    \frac{\int_G 
      \left\langle
        \delta\left(\Phi(k \mathbf{x})\right)\ \left|\nabla_{\mathbf{x}} 
          \Phi(k \mathbf{x}) \right| \right\rangle d^s x}{ k \int_G d^s x}\\
    =&
    \frac{\int_{kG} \left\langle
        \delta\left(\Phi(\mathbf{R})\right)\ \left|\nabla_{\mathbf{R}} 
          \Phi(\mathbf{R}) \right| \right\rangle d^s R}{\int_{kG} d^s R}\\
    =&
     \frac{\int_{kG} \rho(\mathbf{R}) d^s R}{\int_{kG} d^s R}
  \end{split}
\end{equation}
where $kG=\{k \mathbf{x} : \mathbf{x} \in G\}$ is the rescaled region
and we have introduced the (expected) nodal density
\begin{equation}
  \rho(\mathbf{R})=\left\langle
    \delta\left(\Phi(\mathbf{R})\right)\ \left|\nabla_{\mathbf{R}} 
      \Phi(\mathbf{R}) \right| \right\rangle  \ .
\end{equation}
Note that the random wave model is translation invariant with respect
to translations parallel to the boundary $R_s=0$ such that the 
nodal density only depend on the distance $R_s$ from the boundary
\begin{equation}
  \rho(\mathbf{R})=\rho(R_s)\ .
\end{equation}
As $N\rightarrow \infty$ any probability distribution the random wave model 
becomes a Gaussian process such that any probability distribution
involving $\Phi$ and its derivatives 
$\partial_i \Phi\equiv \frac{\partial \Phi}{\partial R_i}$ at a point $\mathbf R$  is a multivariate Gaussian. 
For the present purpose we need 
$P(\Phi,\partial_1 \Phi,\ldots,\partial_s \Phi)$, 
and more specifically in $P(\Phi=0,\partial_1 \Phi,\ldots,\partial_s \Phi)$. 

The relevant variances and cross-correlations can be calculated 
from the known two-point correlator of the standard random wave model which is given by
\begin{equation}
 \langle u(\mathbf{R_1}) u(\mathbf{R_2}) \rangle
 = 2^{\frac{s-2}{2}}\, \mathrm{\Gamma}\!\!\left(\frac{s}{2}\right)
 \frac{\mathrm{J}_{\frac{s-2}{2}}(|\mathbf{R}_1-\mathbf{R}_2|)}{|\mathbf{R}_1-\mathbf{R}_2|^{\frac{s-2}{2}}}\ .
\end{equation}
The lengthy but straight forward calculation of the nodal density of
the boundary-adapted random wave model
can be performed by generalising Berry's 2-dimensional calculation
\cite{berry-nod}. We refer to \cite{Stelios}
for details of the calculation which leads to the nodal density
\begin{equation}
  \rho(R_s)= \rho_{\mathrm{bulk}}
  \sqrt{\frac{s \ D_{R_1}}{B}} \: \mathrm{F} \! \left( -\frac{1}{2}\, ,
    \frac{1}{2}\, ; \frac{s}{2}\, ; M \right)
  \label{nodaldensity}
\end{equation}
where
\begin{equation}
  \rho_{\mathrm{bulk}} =\left\langle \delta\left( u(\mathbf{R})\right)
    \left|\nabla_{\mathbf{R}} u(\mathbf{R})\right|
  \right\rangle=
  \frac{ \Gamma \! \left( \frac{s+1}{2} \right) }{ \sqrt{s \pi} \:
    \Gamma \! \left( \frac{s}{2} \right) }
  \label{rhobulk}
\end{equation}
is the constant nodal density of the standard random wave model
without boundary
(below we will show that $\rho(R_s) \to \rho_{\mathrm{bulk}}$ as
$R_s\to \infty$) and
$\mathrm{F} \! \left( a, b; c; x
  \right)\equiv  {}_2\mathrm{F}_1 \! \left(a,b;c;x \right) $ 
is a hypergeometric function \cite{abram}. The nodal density
\eqref{nodaldensity}
depends on the distance $R_s$ from the boundary via the covariance functions
\begin{align}
  B \equiv& \langle \Phi(\mathbf{R})^2 \rangle  &=& 1 - \mathrm{\Gamma}\!\!\left(\frac{s}{2}\right) \frac{\mathrm{J}_{\frac{s-2}{2}}(2R_s)}{R_s^{\frac{s-2}{2}}}\\
  D_{R_1} \equiv& \langle \partial_1 \Phi (\mathbf{R})^2 \rangle &=&
  \frac{1}{s} - \frac{\mathrm{\Gamma}\!\!\left(\frac{s}{2}\right)}{2} 
  \frac{\mathrm{J}_{\frac{s}{2}}(2R_s)}{R_s^{\frac{s}{2}}} \\
  D_{R_s} \equiv& \langle \partial_s \Phi(\mathbf{R})^2 \rangle &= & 
  \frac{1}{s} + \mathrm{\Gamma}\!\!\left(\frac{s}{2}\right)
  \frac{\mathrm{J}_{\frac{s}{2}}(2R_s) -2 R_s\mathrm{J}_{\frac{s+2}{2}}(2R_s)}{2 R_s^{\frac{s}{2}}} \\
  K \equiv& \langle \Phi(\mathbf{R}) \partial_s
  \Phi(\mathbf{R})\rangle &=&
  \mathrm{\Gamma}\!\!\left(\frac{s}{2}\right)
  \frac{\mathrm{J}_{\frac{s}{2}}(2R_s)}{R_s^{\frac{s-2}{2}}}
\end{align}
where $J_n(x)$ is the $n$-th Bessel function
and the abbreviation
\begin{equation}
  M \equiv 1 - \frac{BD_{R_s} - K^2}{B D_{R_1}}\ .
\end{equation}
Let us now consider the asymptotic behaviour of the nodal density
\eqref{nodaldensity} as $R_s \to \infty$. Using the known 
asymptotic behaviour of Bessel functions  and for 
the hypergeometric function one then obtains
$\rho(R_s) \to \rho_{\mathrm{bulk}}$  with the leading order smooth
and oscillatory corrections given by
(see \cite{Stelios} for details of the calculation)
\begin{equation}
  \frac{\rho(R_s)}{\rho_{\mathrm{bulk}}} = 
    1  +  C^{\mathrm{sm}}_s  R_s^{-(s-1)} + C^{\mathrm{osc}}_s  \frac{\cos(2R_s - \frac{s-1}{4} \, \pi)} {R_s^{\frac{s-1}{2}}}   
    + \mathit{O}(R_s^{-(s-1)}) 
    \label{nodal_asy}
\end{equation}
where
\begin{align}
C^{\mathrm{sm}}_s =& -
\frac{(s-1)\Gamma(s)\Gamma\left(\frac{s}{2}\right)}{2^{s+2}\sqrt{\pi}(s+2)
\Gamma\left(\frac{s+1}{2}\right)}
\\
C^{\mathrm{osc}}_s =&
\frac{\Gamma \left(\frac{s}{2}\right)}{\sqrt{\pi}} \ .
\end{align}
The oscillatory part in \eqref{nodal_asy} decays much slower
than the smooth corrections. For nodal volumes one needs to integrate
and we will see that the smooth correction will dominate over the
oscillatory part. Note that \eqref{nodal_asy}  has oscillatory terms of order
$\mathit{O}(R_s^{-(s-1)})$ which are formally of the same order as
the smooth correction.\\ 
On the boundary $R_s=0$ one finds
\begin{equation}
  \frac{\rho(0)}{\rho_{\mathrm{bulk}}} = \frac{1}{2} \frac{ \sqrt{s}
    (s-1)\, \Gamma \! \left(\frac{s}{2}\right)^2} {\sqrt{s+2} \,\,
    \Gamma \! \left(\frac{s+1}{2}\right)^2} < 1
\end{equation}
which is consistent with the expectation that a Dirichlet boundary
condition will lead to a suppression of the nodal density near the
boundary. 

We can now come back to the estimate of the rescaled nodal
volume inside a given bounded region. For the standard random wave
model the rescaled nodal volume is equal to the 
constant nodal density $\left\langle \sigma_G \right\rangle =
\rho_{\mathrm{bulk}}$. For the boundary adapted random wave
model there are corrections which become stronger close to the
boundary. To be specific let $G$ be a cylindrical region in
$\mathbbm{R}^s$ of height $a$ such that the bottom is a 
connected bounded $s-1$-dimensional region in the hyperplane
$x_s=0$. The volume of $G$ is $\mathcal{V}=a \mathcal{S}$ where
$\mathcal{S}$ is the $s-1$-dimensional hypervolume of the bottom.
The rescaled nodal volume is
\begin{equation}
  \left\langle \sigma_G \right\rangle= \frac{\mathcal{S} \int_0^{ka}
    \rho(R_s) dR_s}{\mathcal{V} k}
  =
  \rho_{\mathrm{bulk}}\left(1  + \frac{\mathcal{S}}{\mathcal{V} k} \int_0^{ka}
    \frac{\rho(R_s) - \rho_{\mathrm{bulk}}}{\rho_{\mathrm{bulk}}} d
    R_s\right) \ .
  \label{sigma_rw}
\end{equation}
In the high energy limit $E=k^2 \to \infty$ the rescaled nodal density
converges to $\rho_{\mathrm{bulk}}$. Indeed, 
$\rho(R_s)$ is a bounded function that converges to
$\rho_{\mathrm{bulk}}$ as $R_s \to \infty$ such that the integral
$\int_0^{ka}  \frac{\rho(R_s) - \rho_{\mathrm{bulk}}}{\rho_{\mathrm{bulk}}} d
R_s $ is at most of order $\mathit{o}(k)$ and the correction term
is at most  $\mathit{o}(1)$ for $k\to \infty$. 
The form of the leading order correction can be obtained 
from the asymptotic expansion of the nodal density \eqref{nodal_asy}.
In the $2$-dimensional case one finds 
\begin{align}
  \int_0^{ka}  \frac{\rho(R_s) - \rho_{\mathrm{bulk}}}{\rho_{\mathrm{bulk}}} d
  R_s \sim& -I_2 + C^{\mathrm{sm}}_2  \int_{1}^{ka}  R_2^{-1} dR_2 \\
  \sim& -I_2 + C^{\mathrm{sm}}_2  \log(ka) 
\end{align}
where $I_2= -\int_0^\infty \left(\frac{\rho(R_2)}{\rho_{\mathrm{bulk}}}
  -1 - C^{\mathrm{sm}}_2 \theta(R_2-1)   R_2^{-1}  \right) dR_2$ is a
constant term.
For any higher dimension $s\ge 3$ we may define the constant
\begin{equation}
  I_s= -\int_0^\infty \left(
    \frac{\rho(R_s)}{\rho_{\mathrm{bulk}}}
    -1 \right) dR_s \ .
\end{equation}
Altogether we obtain
\begin{equation}
  \left\langle \sigma\right\rangle_G =
  \begin{cases}
    \rho_{\mathrm{bulk}}\left(1-\frac{\mathcal{S}}{\mathcal{V}}
      \frac{\log k}{ 32 \pi k} +\mathit{O}(1/k)  \right)
    & \text{for $s=2$,}\\
    \rho_{\mathrm{bulk}}\left(1- \frac{\mathcal{S}}{\mathcal{V}}
      \frac{I_s}{ k} +\mathit{O}(1/k^{3/2})  \right)\ .
  \end{cases}
  \label{sigma_asy}
\end{equation}
By numerical integration one obtains the coefficients $I_3\approx
0.758$ and $I_4=0.645$. 

Let us also mention that Berry
has shown that $\langle \sigma_G^2\rangle - \langle \sigma_G \rangle^2= \mathit{O}\left(\frac{\log
k}{k}\right)$  \cite{berry-nod}  in $s=2$ dimensions which implies
that the distribution of nodal volumes $P(\sigma)$ for a finite energy
interval is very narrow  and converges to
\begin{equation}
  P(\sigma)= \delta \left(\sigma-\rho_{\mathrm{bulk}}\right)
  \label{Psigma}
\end{equation}
as $E=k^2\to\infty$. 
The same scaling also applies to a random wave model on the 2-sphere
as shown by Wigman in \cite{wigman-fluct}. The latter work by Wigman
also states that for $s\geq 3$ one may show that the rescaled nodal volume variance for the
$s$-sphere is bounded by $O(k^{-(s-3)})$. This implies that
\eqref{Psigma} applies to all dimensions for random waves on spheres.
As curvature effects should not change these scalings
one expects the limiting
distribution \eqref{Psigma} also in the case of the Euclidean random
waves
with boundary considered here.

At the end of this section let us come back to more general manifolds
with chaotic ray dynamics. Berry has introduced the standard  Gaussian random-wave model
without boundary as a model for wave functions in this case and he also
proposed to include boundary effects with
the boundary-adapted random-wave model. In \cite{berry-nod} he
conjectured that for $s=2$ the $\frac{\log k}{k}$ corrections of the
boundary-adapted random wave model should also apply to chaotic
billiards.
Here we extend his conjecture to arbitrary dimensions, i.e. 
the asymptotic behaviour of the nodal volume is described by
\eqref{sigma_asy} including the leading order correction terms
and no free parameters as $\mathcal{V}$ should be replaced by the
volume
of the manifold $\mathcal{M}$ and $\mathcal{S}$ is the hypervolume of
its boundary. For higher order corrections one may need to include
curvature effects which are not represented in the random wave model.
In the case $s=2$ terms of order
$\mathit{O}\left(k^{-1}\right)$
in the nodal volume of random waves in a cylindrical region
explicitly contain the height $a$ which is not well defined when one
tries to translate the result to general billiards \cite{berry-nod}.

\section{Signatures of wave chaos and integrability}
\label{discussion}

In the previous two chapters we have given a detailed account
of nodal volume statistics for two paradigmatic systems in
arbitrary dimension. The $s$-dimensional cuboid is a paradigm for
a regular shape with separable wave functions while planar random waves
are a model for wave functions on irregular shapes with chaotic ray dynamics.
In both cases we have found limiting distributions for the rescaled
nodal volume. The limiting distributions in the two paradigms have a very different character:
for random waves we find a delta-function while the cuboid's limiting 
distribution has a finite support. It is consistent with Berry's random wave
conjecture that the distribution for all irregular shapes will be a delta 
function at $\sigma=\rho_\mathrm{bulk}$ (see \eqref{rhobulk}) which only depends on the
dimension and nothing else.
For regular shapes other than the cuboid we may refer to the 
analogous calculations for
nodal count distributions \cite{BGS,GL} which showed that many features such as
the types of singularities near the upper and lower end of the support
are universal. We conjecture that the same is true for the nodal
volume distributions studied here though the actual values for the upper and
lower end of the support may be system dependent.
One clear signature of irregular versus regular shapes is the behaviour
of the fluctuations of rescaled nodal volumes as the energy increases.
For irregular shapes one expects that the variance decreases while it remains
finite and bounded for regular shapes.\\
Another interesting difference 
between regular and irregular shapes in dimension $2$ is the different
nature of the boundary correction to expected rescaled nodal volumes --
these go like $1/k$ for the cuboid (and, conjecturally, for other regular shapes with separable wave functions) 
and like $\log k/k$ for the random wave 
model. It has been shown by numerical computations that the 
boundary adapted random wave model 
gives an accurate account of the nodal density near a boundary 
\cite{berry-nod3} and it was conjectured by Berry that this may be seen in
any chaotic billiard.\\
Let us also note, that the average rescaled nodal volume 
for the cuboid is always larger than the one for a random wave.
This is expected as the nodal surfaces of a separable function
generally intersect in a checker-board structure while nodal
intersections are avoided in random waves \cite{Monastra} --
this effective repulsion of nodal surfaces leads to the
decrease in the expected nodal volume. Again, the same 
decrease can be 
expected for eigenfunctions in irregular shapes when compared to
separable eigenfunctions of a regular shape.

\end{document}